\documentclass[12pt, a4paper]{article}

\usepackage{amsthm, amsfonts, amssymb, braket, framed, graphicx, fancyhdr, eurosym, url, multirow}
\usepackage{amsmath}
\usepackage{subcaption}
\usepackage[top=30mm, bottom=30mm, right=25mm, left=25mm, headheight=15.72pt]{geometry}
\usepackage[utf8]{inputenc}
\usepackage[british,UKenglish,USenglish,english,american]{babel}
\usepackage{mathabx}
\usepackage{mathtools}

\usepackage{cite}
\usepackage{hyperref}
\usepackage{upgreek}
\usepackage{comment}
\usepackage{mathrsfs}
\usepackage[dvipsnames]{xcolor}
\usepackage{tikz}
\usetikzlibrary{arrows,intersections,decorations.markings,}

\usepackage[T1]{fontenc}
\numberwithin{equation}{section}

\begin{document}

\begin{titlepage}

\begin{center}

\parbox[t]{\textwidth}{\centering \LARGE	\bf
	\vspace{1em}
	Are there Goldstone bosons in ${d\leq z+1}$\ ? 
}

\vspace{3em}

{\large 
		Riccardo Argurio, Daniel Naegels and Antoine Pasternak}

\vspace{2em}
\itshape\centering
{Physique Th\'eorique et Math\'ematique and International Solvay Institutes\\ Universit\'e Libre de Bruxelles, C.P. 231, 1050 Brussels, Belgium.}

\end{center}

\vspace{3em}

\noindent
\begin{center}
\bf \small Abstract
\end{center}

\noindent
We study the viability of spontaneous breaking of continuous symmetries in  theories with Lifshitz scaling, according to the number of space-time dimensions $d$ and the dynamical scaling $z$. 
Then, the answer to the question in the title is no (quantum field theoretically) and yes (holographically).
With field theory tools, we show that symmetry breaking is indeed prevented by large quantum fluctuations when $d\leq z+1$, as expected from scaling arguments.
With holographic tools, on the other hand, we find nothing that prevents the existence of a vacuum expectation value.
This difference is made possible by the large $N$ limit of holography. An important subtlety in this last framework is that in order to get a proper description of a conserved current, renormalization of the temporal mode of the bulk vector requires an alternative quantization.
We also comment on the implications of turning on temperature. 

\end{titlepage}

\tableofcontents


\section{Introduction}
Spontaneous symmetry breaking is known to be fragile in situations where fluctuations are large. This is true for thermal fluctuations in two spatial dimensions \cite{Mermin:1966fe,Hohenberg:1967zz}, and for quantum fluctuations in two relativistic space-time dimensions \cite{Coleman:1973ci}. The latter result is quoted as saying that there are no Goldstone bosons in two dimensions. In fact, it is precisely the large quantum fluctuations of the would-be Goldstone boson that destroy any vacuum expectation value giving a symmetry breaking order. 

There is however a class of theories which still displays spontaneous symmetry breaking in such situations: theories with a large number $N$ of constituents are known to have ordered phases, in the $N\to\infty$ limit \cite{Coleman:1974jh,Gross:1974jv}. It can be seen that the large quantum fluctuations are actually suppressed by a $1/N$ power \cite{Witten:1978qu}. This is precisely the case for theories which have a holographic dual. 

It was shown in \cite{Argurio:2016xih} that indeed $AdS_3$ holography allows for spontaneous symmetry breaking in its dual two-dimensional QFT. It was realized there that, however, the holographic set-up keeps a score of the peculiarity related to being in two dimensions: the bulk vector dual to the current of the symmetry which was to be broken, has to undergo alternative quantization to properly account for the conserved current Ward identities.

In the present paper, we wish to extend the discussion of the survival of spontaneous symmetry breaking to theories with Lifshitz scaling. It is expected that below a certain space-time dimension $d$, as a function of the scaling $z$ of the time direction, again the quantum fluctuations will be strong enough to destroy the symmetry breaking order (see for instance \cite{Griffin:2013dfa}). Below, we confirm that indeed (in theories preserving time-reversal invariance) for $d\leq z+1$, one-point functions are set to zero by large quantum fluctuations of the would-be Goldstone boson.
Interestingly, in contrast to relativistic QFTs where only two dimensions are singled out as a particular case, in Lifshitz QFTs there is a whole range of dimensions in which spontaneous symmetry breaking is in principle forbidden.

We then explore what happens to spontaneous symmetry breaking in Lifshitz scaling theories which are described holographically (see \cite{Kachru:2008yh,Taylor:2008tg,Baggio:2011cp,Ross:2011gu,Korovin:2013bua,Chemissany:2014xsa,Taylor:2015glc,Argurio:2017irz}), i.e.~in a large $N$ limit. We extend the analysis of \cite{Argurio:2017irz} to the above mentioned case of $d\leq z+1$. As in the relativistic case \cite{Argurio:2016xih}, we find that alternative quantization for the vector is needed, albeit in the present case of Lifshitz scaling, only the temporal component of the vector is involved. This quantization which treats differently space and time is actually needed to enforce the proper gauge invariance of the generating functional.

The paper is organized as follows. In section \ref{secQFT} we consider a QFT invariant under the Lifshitz group,  time-reversal and a global $U(1)$ symmetry. We discuss under which conditions a one-point function vacuum expectation value survives quantum corrections. We find the condition to be $d>z+1$, in agreement with a naive dimensional argument. We then proceed in section \ref{secHOLO} to analyze an equivalent holographic set-up. With the usual artillery of holographic renormalization, we establish which counterterms need to be selected in order to obtain the correct gauge invariance of the generating functional, and hence reproduce the usual Ward identities for the conserved current. Such counterterms impose alternative quantization for the temporal component of the bulk vector, i.e.~the leading term in the near boundary expansion is identified with the VEV rather than the source. Finally, in section \ref{secDISC} we comment on a few open questions.

\section{Quantum corrections to the symmetry breaking VEV}
\label{secQFT}
We present in this section a generalisation of the argument by Coleman \cite{Coleman:1973ci} for quantum field theories with Lifshitz scaling. As a reminder, Coleman's theorem states that for a relativistic theory in two dimensional space-time, at the quantum level, there cannot be any spontaneous breaking of symmetries that would lead to Goldstone bosons. The idea behind this argument is that for this specific space-time dimension, massless scalars are ill-defined and so is the ``would-be'' Goldstone boson associated to the symmetry breaking. Physically, the interpretation is that quantum fluctuations are large enough to erase any notion of order, leading to the impossibility of having spontaneously broken symmetries.

The different Lifshitz theories being studied are identified by the number of space-time dimensions $d$ and the value of the dynamical critical exponent $z$. 
The argument is built with respect to a general action of the Lifshitz type\footnote{Lifshitz symmetry is considered here as not being emergent. Namely, \eqref{fundatmentalTheory} is seen as defining a fundamental theory.} invariant under a global continuous symmetry group. For simplicity, we consider the theory of a complex scalar $\psi$ that is charged under a $U(1)$ global symmetry, invariant under time-reversal\footnote{We will comment in section \ref{secDISC} the more general situation of a theory with broken time reversal invariance.} and that possesses a potential $V$ depending only on the modulus of $\psi$. To trigger the spontaneous symmetry breaking at the classical level, we suppose that $V$ is minimal around a vacuum expectation value $v$ for $|\psi|$,  and there it takes the value zero for simplicity. The action is then given by
\begin{eqnarray}
S\left[\psi\right] &=& \int  \mathrm{d} t \mathrm{d}^{d-1} x  \, \left( \partial_{t}\psi\partial_{t}\psi^\ast - (-1)^{z}\xi^{2}\psi\nabla^{2z}\psi^\ast - V(\psi\psi^\ast )\right)
\label{fundatmentalTheory}
\end{eqnarray}
where $z$ is the dynamical critical exponent (as we motivate later, we can take $z\geq 1$), $\xi$ is a positive real number without dimensions and $d\geqslant 2$ (to discuss Lifshitz scaling we need at least one spatial direction and one time direction). We note that $\psi$ has dimension
\begin{equation}\label{dimphy}
[\psi]=\frac{d-1-z}{2}\ .
\end{equation} 

Doing a perturbation around the classical VEV, the physical field can be written as
\begin{eqnarray}
\psi(x)\equiv \left(v+\sigma\left(x\right)\right)e^{i\theta(x)}
\end{eqnarray}
where $\sigma$ and $\theta$ are small fluctuations. The phase-field $\theta$ corresponds to the longitudinal direction of the action of $U(1)$ on the physical field, hence, it corresponds to the Goldstone boson if spontaneous symmetry breaking is allowed. Since we perform an analysis till the quadratic order (small perturbations), the dynamics of $\theta$ is dictated by the free effective action
\begin{equation}
S\left[\theta \right]=\int  \mathrm{d} t \mathrm{d}^{d-1} x \, \, v^2 \left( \partial_{t}\theta\partial_{t}\theta - (-1)^{z}\xi^{2}\theta\nabla^{2z}\theta - \xi^2 \lambda^{2z} \theta^2 \right)\ .
\end{equation}
A mass term for $\theta$ with parameter $\lambda$ is added by hand in order to confront the cases of spontaneous and explicit symmetry breaking. This parameter can also be viewed as an infrared regulator.

All we need for our argument is the (non-time-ordered) two-point function of $\theta$
\begin{eqnarray}
\left.\left< \theta(t,\vec{x})\theta(0) \right>\right\rvert_{\lambda} & = & \frac{\pi}{(2\pi)^d  \xi v^2}\int \mathrm{d}^{d-1} p \, \frac{e^{i\vec{p}\cdot\vec{x}-i\xi \sqrt{p^{2z} + \lambda^{2z}} t}}{\sqrt{p^{2z} + \lambda^{2z}}}\ . \label{2points0}
\end{eqnarray}
where $p \equiv \parallel \! \vec{p} \!\parallel$. On purely dimensional grounds, the behavior at large (spatial) separation of the propagator for $\theta$ is dictated, in the massless limit, by the dimension of $\psi$, \eqref{dimphy}. We thus expect the correlations to vanish at large separations only for positive dimensions, i.e.~for $d>z+1$. Conversely, for $d\leq z+1$, we expect large long range correlations that can potentially spoil any vacuum expectation value.

We are now going to show that indeed, after renormalization, the VEV is preserved in the former case, and is set to zero in the latter. We will follow an argument similar to the one given in \cite{Ma:1974tp} for the relativistic case, which is essentially equivalent to computing the one-loop correction to the $\psi$-tadpole.

First of all, if $\theta$ is approximated by a free field, we can write $\theta \equiv \theta^+  + \theta^-$ where $\theta^+$ is associated to the positive energy modes and is proportional to an annihilation operator, $\theta^-$ is associated to the negatives energy modes and is proportional to a creation operator. If we consider the two-point function of $\theta$, we find
\begin{equation}
\left< \theta(x) \theta(0)\right> = \left< \theta^+\!(x) \theta^-\!(0) \right>  = \left< \left[\theta^+\!(x) ,  \theta^-\!(0)\right] \right> .
\end{equation}
We now evaluate the one-point function of $\psi$ using its decomposition in terms of the fluctuations $\sigma$ and $\theta$
\begin{equation}
\left< \psi(x) \right> = v \left< e^{i\theta(x)}\right> =v \left< e^{i \theta^-\!(x) }e^{ i \theta^+\!(x)}e^{1/2\left[ \theta^-\!(x) , \theta^+\!(x) \right] }\right> = v \, e^{-1/2\left< \left[ \theta^+\!(x) , \theta^-\!(x) \right]\right> }= v \, e^{-1/2\left<  \theta(0) \theta(0) \right> } , \label{1pointNR}
\end{equation}
where we used, besides the previous arguments, also the fact that $\sigma$ is a  massive perturbation around $v$. We thus see that in order to certify whether the VEV is maintained at the quantum level, we need the two-point function for $\theta$ at vanishing distance in time and space.

Obviously, such a limit $(t,\vec{x})\rightarrow 0$ can lead to a UV divergence, naively giving an ill-defined one-point function above. However it is known how to deal with such divergence through renormalization. In order to disentangle potential IR divergences, we use the theory regulated by the small mass $\lambda$, 
guided by the expectation that explicit symmetry breaking is always viable and a non-zero value for the order parameter should be found in that case. In consequence, the limit $\lambda \rightarrow 0$ alone must have something to tell us about the possibility of spontaneous symmetry breaking. 

We now use (\ref{2points0}) evaluated at coinciding points to find the needed expression. Analogously, this computation can be seen as the evaluation of the one-loop correction to the tadpole. Following standard manipulations (see e.g.~\cite{Porrati:2016lzr} or \cite{Keranen:2016ija} for a similar context), we have
\begin{eqnarray}
\left.\left< \theta(0)\theta(0) \right>\right\rvert_{\lambda} & = &  \frac{\pi}{(2\pi)^d  \xi v^2}\int \mathrm{d}^{d-1} p \, \frac{1}{\sqrt{p^{2z} + \lambda^{2z}}} \nonumber \\
& = &  \frac{\Gamma\left( (d-1)/2z\right)\, \Gamma\left( (z+1-d)/2z\right) }{(4\pi)^{d/2}\Gamma((d-1)/2) \, z \,  \xi v^2} \lambda^{d-1-z} \ .\label{2pointsL}
\end{eqnarray}
From the integral, we note that the IR behavior will be dictated by the presence of $\lambda$, while a UV divergence might appear when $d\geq z+1$. This latter divergence ends up being encoded in the Gamma function $\Gamma\left( (z+1-d)/2z\right)$. The IR behavior will give a vanishing result for $d>z+1$, and a diverging one for $d<z+1$. At the same time, the Gamma function is always regular for $d<z+1$, while it can have singularities for $d\geq z+1$ (more specifically, it diverges for $d=z+1+2nz$, with $n$ a positive or null integer). The limiting case is obviously $d=z+1$, actually the only one where we need to disentangle UV and IR divergences.

For $d=z+1$, let us treat this case with dimensional regularization. Setting $d \rightarrow z+1-2z\epsilon$ gives first
\begin{eqnarray}
\left.\left< \theta(0)\theta(0) \right>\right\rvert_{\lambda}^\epsilon & = & \frac{\Gamma\left(1/2 \right) }{(4\pi)^{(z+1)/2}\Gamma(z/2) \, z \,  \xi v^2} \left( \epsilon^{-1} + \text{const.} -2z \ln\lambda + O(\epsilon) \right)\ .
\end{eqnarray}
We obtain a UV-regular expression keeping only the finite $\lambda$-dependent piece (and introducing for dimensional reasons the renormalization scale $\mu$):
\begin{eqnarray}
\left.\left< \theta(0)\theta(0) \right>\right\rvert_{\lambda}^{\mathcal{R}_{\text{UV}}} & \equiv & \lim_{\hspace{0.2cm}\epsilon \rightarrow 0^+}\left( \left.\left< \theta(0)\theta(0) \right>\right\rvert_{\lambda}^\epsilon  -  \left.\left< \theta(0)\theta(0) \right>\right\rvert_{\mu}^\epsilon   \right) \nonumber \\
&=& - \frac{\Gamma\left(1/2 \right) }{(4\pi)^{(z+1)/2}\Gamma(z/2)  \,  \xi v^2}\ln\left(\lambda/\mu \right)^2\ .
\end{eqnarray}
This expression is free from UV divergence thanks to renormalization, but still has an IR divergence when $\lambda\to 0$. We can thus conclude that the massless same-point correlator diverges to $+\infty$ when $d=z+1$.

For $d<z+1$, we see from \eqref{2pointsL} that in the limit $\lambda\to 0$ the expression also diverges to $+\infty$ (recall we assume $d\geq2$), without any need to regularize and renormalize in the UV.

For $d>z+1$, we would need to regularize and renormalize in certain cases as discussed above. However, we see in \eqref{2pointsL} that the result is multiplied by a positive power of $\lambda$, which will always win in the $\lambda\to 0$ limit against any term involving $\ln \lambda$. We thus conclude that the correlator in this case always vanishes in the massless limit.

Now, going back to the expression \eqref{1pointNR}, inserting the UV-renormalized two-point function, we observe that the VEV is preserved when $d>z+1$ while it is set to zero when $d\leq z+1$. We summarize the results in the table below.

\begin{center}
\begin{tabular}{c||c|c|c}
		\begin{tabular}{c}
		Condition \\
		for $d$ and $z$
		\end{tabular} &
		\begin{tabular}{c}
		$ \lim_{\lambda \rightarrow 0}$\\
		$\left.\left< \theta(0)\theta(0) \right>\right\rvert_{\lambda}^{\mathcal{R}_{\text{UV}}} $
		\end{tabular} &
		\begin{tabular}{c}
		$\left< \psi(x) \right>^{\mathcal{R}_{\text{UV}}} $
		\end{tabular} &
		\begin{tabular}{c}
		Spontaneous \\ symmetry breaking
		\end{tabular}
		  \\
		\hline\hline
		$d > z+1$ & 0 & $v$ & yes\\
		\hline
		$d \leqslant z+1$ & $+\infty$ & 0 & no
	\end{tabular}
\end{center}

We have thus generalized the Coleman theorem on the possibility of having spontaneous symmetry breaking to Lifshitz theories. The argument  is essentally based on contradiction. By considering a generic $U(1)$ theory presenting Lifshitz scaling symmetry, the hypothesis that $U(1)$ is spontaneously broken leads to the presence of a massless field, the would-be Goldstone boson. We then observed that for $d\leq z+1$ the latter is not well-defined, leading to large quantum fluctuations that set the VEV to zero. Hence no spontaneous symmetry breaking can occur in those dimensions.

We now turn to discuss the same kind of theory, but with a large $N$ number of constituents. We employ holography to study it, and enquire whether the large $N$ limit can restore an ordered vacuum.

\section{Holographic renormalization and symmetry breaking in $\boldsymbol{d\leq z+1}$}
\label{secHOLO}
In this section we consider a theory with the exact same symmetry properties, but from a holographic perspective. This is tantamount to say that the QFT under consideration, besides being in the large $N$ limit, is also generically strongly coupled. We will use a set-up in all similar to the one considered in \cite{Argurio:2017irz}, though we will implement time-reversal symmetry to be consistent with the discussion in the previous section.

On the bulk, gravity side of the holographic correspondence, we thus introduce a complex scalar $\phi$ charged under a $U(1)$ gauge symmetry. The charge is set to unity and the corresponding gauge field is $A$. To reproduce a QFT invariant under Lifshitz scaling, this matter content has to live on a curved space-time in $d+1$ dimensions dominated by the presence of a background massive vector field $B$ \cite{Taylor:2008tg}. If it is defined as\footnote{Besides the obvious requirement of keeping $\beta$ real, the condition $z\geq 1$ has strong physical motivations, both in QFT and in holography \cite{Hoyos:2010at}, essentially $z<1$ would lead to causality violations.}
\begin{eqnarray}
B \equiv \frac{\beta}{r^z}\mathrm{d}t \hspace{0.5cm} & \text{with} & \hspace{0.5cm} \beta \equiv \sqrt{\frac{2(z-1)}{z}}, \, \, \,z \geq 1 \ ,
\end{eqnarray} 
then the background metric reads
\begin{equation}
\mathrm{d}s^2 \equiv g_{mn}\mathrm{d}x^m \mathrm{d}x^n = \frac{\mathrm{d}r^2}{r^2} - \frac{\mathrm{d}t^2}{r^{2z}} + \frac{\mathrm{d}x^2_j}{r^2} \ ,
\end{equation}
with $j$ running from 1 to $d-1$, and is isometric under a Lifshitz scaling and the rotations of space coordinates. The part of this metric that is orthogonal to $B$ is given by
\begin{eqnarray}
\gamma_{mn} \equiv g_{mn} + \beta^{-2}B_m B_n \hspace{0.5cm} & \text{so that} & \hspace{0.5cm} \gamma_{mn}B^n = 0.
\end{eqnarray} 
A general action invariant under the Lifshitz symmetry group and time-reversal for a scalar $\phi$ and a massless vector $A$ is then given by
\begin{eqnarray}
S\left[A_m,\phi\right] &=& \int \mathrm{d}^{d+1}x \,\, \sqrt{-g} \,\, \Bigg\lbrace -\frac{1}{4} \gamma^{mn}\left( \gamma^{pq} - \frac{2\kappa}{\beta^2}B^p B^q\right) F_{mp}F_{nq}   \nonumber \\
& &   - \left( \gamma^{mn} - \frac{1}{c^2 \beta^2}B^m B^n\right) \left(D_m \phi\right)^\ast D_n \phi  - m^2 \phi^\ast\phi  \Bigg\rbrace . \label{S}
\end{eqnarray}
where $F_{mn} = \partial_m A_n - \partial_n A_m$ and $D_m = \partial_m  - i A_m$ as usual. Since we will not consider correlators involving the QFT stress-energy tensor complex, $B$ as well as the metric $g$ are meant as non-dynamical fields. Similarly, we will neglect backreaction of the scalar on them. This theory has three free parameters : $\kappa$, $c^2$ and $m^2$.

We list here the equations of motion that are obtained from the action above 
\begin{eqnarray}
 \partial_m \left(  \frac{\sqrt{-g}}{2} \left( \gamma^{mn}\left( \gamma^{pq} - \frac{2\kappa}{\beta^2}B^p B^q\right) - \gamma^{pn}\left( \gamma^{mq} - \frac{2\kappa}{\beta^2}B^m B^q\right)\right)  F_{nq} \right) & & \nonumber \\
   - i \sqrt{-g} \left( \gamma^{pq} - \frac{1}{c^2 \beta^2}B^p B^q\right)\left( \phi^\ast D_q\phi - \phi\left(D_q \phi\right)^\ast \right) &=& 0\ , \label{eqV}\\
   D_m\left( \sqrt{-g} \left(\gamma^{mn} - \frac{1}{c^2\beta^2} B^m B^n\right) D_n \phi  \right)   - \sqrt{-g}m^2\phi &=& 0\ . \label{eqP}
\end{eqnarray}
Of course, when taking the variation of the action with respect to the dynamical degrees of freedom, one has to pay attention to the boundary terms that will play a prominent role in the holographic renormalization.

Since the radial mode of the vector $A$ does not source any operator on the QFT side of the correspondence, we can partially fix the gauge freedom by putting it to zero (i.e.~we work in the radial, or holographic, gauge)
\begin{eqnarray}
A_r &=& 0\ . \label{gauge}
\end{eqnarray}
The spatial modes can be split into transverse and longitudinal modes
\begin{eqnarray}
A_i \equiv T_i +\partial_i L  \hspace{0.5cm} & \text{with the condition} &\hspace{0.5cm} \partial_i T_i = 0\ .
\end{eqnarray}
Finally, we consider the real and imaginary parts of the scalar separately 
\begin{eqnarray}
\phi & \equiv & \frac{\rho + i \pi }{\sqrt{2}}\ .
\end{eqnarray}

The gauge transformations in their infinitesimal form for the newly introduced fields read
\begin{equation}
\delta _\alpha \rho  =  - \alpha \pi\ , \qquad
\delta _\alpha \pi  =  + \alpha \rho\ , \qquad
\delta _\alpha A_t  =  \partial_t\alpha\ , \qquad
\delta _\alpha L  =  \alpha \ ,
\end{equation}
where $\alpha$ is now a function of $t$ and $\vec{x}$ only (to preserve the holographic gauge). All other quantities are gauge invariants.

We now want to switch on a background for the scalar to enforce the symmetry breaking in the QFT. So, we introduce $\phi_B$ that only depends on the $r$ coordinate, and shift
\begin{eqnarray}
\rho & \rightarrow & \phi_B + \rho. 
\end{eqnarray}
Moreover, we prescribe that all the degrees of freedom that we described in the previous section are small fluctuations on top of this background. Assuming the gauge parameter is similarly small, gauge transformations now read 
\begin{equation}
\delta _\alpha \rho  =  - \alpha \pi \approx  0\ , \qquad
\delta _\alpha \pi  =  + \alpha (\phi_B +\rho)   \approx  \alpha\phi_B\ ,\qquad
\delta _\alpha A_t  =  \partial_t\alpha\ , \qquad
\delta _\alpha L  =  \alpha \ .
\end{equation}

First, we find the equation for the background from (\ref{eqP}) :
\begin{eqnarray}
r\partial_r \left( r \partial_r \phi_B \right) - (d+z-1) r \partial_r \phi_B - m^2 \phi_B  &=& 0 \label{eqPB}.
\end{eqnarray}
The gauge fixing we performed in (\ref{gauge}) gives us, taking $p = r$ in (\ref{eqV}), the constraint
\begin{eqnarray}
- \kappa r^{2z}\partial_r\partial_t A_t + r^2 \partial_j^2 \partial_r L - \phi_B \partial_r \pi  + \pi \partial_r \phi_B & = & 0.  \label{eqC} 
\end{eqnarray}
Taking  $p = t$ and $p = j$ in (\ref{eqV}) gives the equations for the temporal and spatial modes of the vector. We also apply the projectors $\left( \delta _{ij} \partial_k^2 - \delta _{ik}  \partial_k\partial_j \right)/\partial_k^2$ and $\partial_j / \partial_k^2$ on the $p=j$ equation to separate equations for the transverse and longitudinal modes. The real and imaginary parts of (\ref{eqP}) give rise to equations for the real and imaginary parts of the scalar respectively.
All in all, the equations of motion for the dynamical degrees of freedom are
\begin{eqnarray}
r \partial_r \left( r \partial_r A_t \right) - ( d-z-1) r \partial_r A_t + r^2 \partial_j^2\left(  A_t - \partial_t L \right) 
+ \frac{1}{\kappa c^2} \left( \phi_B \partial_t \pi - \phi_B^2 A_t  \right)&=& 0 \label{eqAt}\ , \\
r \partial_r \left( r \partial_r T_i \right) - ( d+z-3) r \partial_r T_i -\kappa r^{2z}\partial^2_t T_i +  r^2 \partial_j^2 T_i - \phi_B^2 T_i & = & 0  \label{eqATi}\ , \\
r \partial_r \left( r \partial_r L \right) - ( d+z-3) r \partial_r L -\kappa r^{2z} \partial_t \left(\partial_t L - A_t \right)   + \phi_B \left( \pi  - \phi_B L  \right) & = & 0  \label{eqAL}\ , \\
r \partial_r \left( r \partial_r \rho \right) - ( d+z-1) r \partial_r \rho- \frac{r^{2z}}{c^2}  \partial_t^2 \rho  
+ r^2\partial_j^2 \rho - m^2\rho  & = & 0  \label{eqR}\ , \\
r \partial_r \left( r \partial_r \pi \right) - ( d+z-1) r \partial_r \pi - \frac{r^{2z}}{c^2}  \left( \partial_t^2 \pi   - \phi_B \partial_t A_t \right) \qquad \qquad &&\nonumber \\
+ r^2 \partial_j^2 \left(  \pi - \phi_B L\right)  - m^2\pi  & = & 0 \ . \hspace{1.5cm} \label{eqPi}.
\end{eqnarray}
Since we considered small fluctuations for the dynamical degrees of freedom, those equations are linear in the fields. The degrees of freedom $\rho$ and $T_i$ are both decoupled from the others.

We now turn to the asymptotic expansions of the fields near the boundary. Starting from the background for the scalar field, the exact solution is
\begin{eqnarray}
\phi_B & = & w \, r^{\frac{\tilde{d}}{2} - \nu} + v \,   r^{\frac{\tilde{d}}{2} + \nu}\ , \label{expPB}
\end{eqnarray}
where $w$ and $v$ are  real numbers and we have defined
\begin{equation}
\tilde{d}  \equiv  d+z-1 \qquad \mbox{and} \qquad \nu \equiv \sqrt{\frac{\tilde{d}^2}{4} +  m^2}\ .
\end{equation}
For simplicity we will take $0<\nu<1$.\footnote{Taking $\nu\geq1$ makes the procedure of renormalization more involved. Note also that for $\nu>\tilde{d}/2$, we would need to set $w=0$ in order for the background not to spoil the asymptotic Lifshitz scaling.} For the fluctuations, the
radial behaviour captured in the equations of motion imposes the following expansions. Leaving aside the spatial index $i$ for the mode $T_i$, we get
\begin{eqnarray}
\rho & \overset{r \rightarrow 0 }{\sim} & \rho_0 \,  r^{\tilde{d}/2 - \nu} + \tilde{\rho}_0 \,  r^{\tilde{d}/2+ \nu} + \cdots \label{expR} \\
\pi & \overset{r \rightarrow 0 }{\sim} & \pi_0  \, r^{\tilde{d}/2- \nu} + \tilde{\pi}_0 \, r^{\tilde{d}/2+ \nu} + \cdots \label{expPi} \\
A_t  & \overset{r \rightarrow 0 }{\sim} & \tilde{a}_0  \, r^{-(2z-\tilde{d})} + \cdots + a_0 + \cdots \label{expAt} \\
T & \overset{r \rightarrow 0 }{\sim} & t_{0} + \cdots +  \tilde{t}_{0}  \, r^{\tilde{d} - 2}  + \cdots \label{expAT}  \\
L & \overset{r \rightarrow 0 }{\sim} & l_0 + \cdots +  \tilde{l}_{0}  \, r^{\tilde{d} - 2} + \cdots 
\end{eqnarray}
where all coefficients are fields with a $(t,x_i)$ dependence. 
We have anticipated here the special case where $d\leq z+1$ (i.e.~$2z \geqslant \tilde{d}$), the opposite case was treated in \cite{Argurio:2017irz}. Dots between leading and subleading orders mean that one can find some more terms by adding powers of $r$ two by two, if $\tilde{d}-2>2$ and/or $2z-\tilde{d}>2$. Logarithms should also be taken into account starting from the order $r^0$ in the expansion of $A_t$ if $\tilde{d}-2z$ is even and from $r^{\tilde{d}-2}$ in the expansions of $T_i$ if $\tilde{d}-2$ is even. 
Finally, because of the presence of the background $\phi_B$, and the particular shape of the Lifshitz metric, further powers in the expansions above appear. However it can be checked that they are all subdominant with respect to the ones shown above (provided all our previous assumptions, that is $\nu<1$, $z\geq1$ and $d\geq2$).

Coefficients crowned with a tilde symbol are leading or subleading modes that we do not want to play the role of sources in QFT. For the scalars, it is just a matter of choice (in this case, it identifies $w$ as an explicit symmetry breaking parameter and $v$ as a VEV), while for the gauge field, it is important because only vector modes without tilde symbol transform non-trivially under the gauge group and can actually play the role of sources for a conserved current in QFT.

Indeed, we can determine the gauge transformations for the coefficients. Since $\rho$ does not transform at linear order under the gauged $U(1)$, we have non-trivial rules for coefficients of $\pi$ only
\begin{equation}
\delta _\alpha \pi_0  =   \alpha w\ , \qquad \qquad
\delta _\alpha \tilde\pi_0  =   \alpha v \ .
\end{equation}
For the gauge vector, only two coefficients transform under the gauge transformation
\begin{equation}
\delta _\alpha a_0  =  \partial_t\alpha \ , \qquad \qquad
\delta _\alpha l_0  =  \alpha \ .
\end{equation}
It is important for the following to note that $\tilde{a}_0$ is a gauge invariant quantity and therefore cannot be the source for the temporal part of a conserved current.

Note that in the limiting, relativistic, case where $d=2$ and $z=1$, i.e. $\tilde{d} = 2$, all leading and subleading terms of the vector modes have the same order in $r$ respectively :
\begin{eqnarray}
A_t  & \overset{r \rightarrow 0 }{\sim} & \tilde{a}_0  \, \ln r +  a_0 + \cdots    \\
\tilde{d} =2 \hspace{0.5cm} : \hspace{1cm}T & \overset{r \rightarrow 0 }{\sim} & \tilde{t}_0  \,\ln r +t_{0} + \cdots \hspace{3cm}  \\
L & \overset{r \rightarrow 0 }{\sim} & \tilde{l}_0  \, \ln r  +  l_0 + \cdots  
\end{eqnarray}
This case was already discussed in \cite{Argurio:2016xih} (see also \cite{Marolf:2006nd,Faulkner:2012gt}) so we will keep $\tilde{d}>2$ from now on.

We can now apply the procedure of holographic renormalization \cite{Bianchi:2001de,Bianchi:2001kw}.
Applying the equations of motion in the expression (\ref{S}), we find an action on the boundary. To regularize divergences, we evaluate it on a slice $r=\epsilon$ close to $r=0$. This procedure defines the regularized action :
 \begin{eqnarray}
S_{reg} & \equiv &\underset{r = \epsilon }{\int}\mathrm{d}^dx \, \, \frac{r^{-\tilde{d}}}{2}\Big\lbrace  r^2 T_i \, r \partial_r T_i - r^2 L  \, r \partial_r\partial_j^2 L- \kappa r^{2z} A_t\,  r \partial_r A_t \nonumber \\
 & & \qquad\qquad\qquad +  \phi_B \, r \partial_r \phi_B + 2 \rho\,  r \partial_r \phi_B +  \rho\,  r \partial_r\rho+  \pi \, r \partial_r \pi \Big\rbrace\ .
 \end{eqnarray}
We need to add some counterterms to get rid of the divergences and to see clearly which coefficient of each expansion seen before is a source for the action. To do it properly, we look at the variation  that has to vanish to satisfy the variational principle (note that $\delta S_{reg}$ is not the variation of $S_{reg}$ given above, but the regularized variation of the action (\ref{S}))
 \begin{eqnarray}
\delta  S_{reg} &=& \underset{r=\epsilon}{\int} \mathrm{d}^{d}x \,\, r^{-\tilde{d}}\, \Big\lbrace r^2 \delta  T_i \, r \partial_r  T_i -  r^2 \delta  L \,\partial_j^2 \,  r \partial_r L - \kappa \,  r^{2z} \delta  A_t \, r \partial_r A_t\nonumber \\
& &\qquad\qquad\qquad + \delta \rho\, r\partial_r \left( \phi_b + \rho \right) + \delta  \pi \, r \partial_r \pi \Big\rbrace\ . \label{varSreg}
 \end{eqnarray}
We will now renormalize this expression for the different sectors separately. We anticipate that the sector that will contain all the subtleties is the one of the temporal and longitudinal components of the vector. We start by treating the other sectors.

For the scalar sector, the procedure goes exactly as in \cite{Argurio:2017irz}. We add the counterterm
\begin{eqnarray}
S^\phi_{ct} & \equiv  &\left(  \tilde{d}/2- \nu\right) \underset{r=\epsilon}{\int} \mathrm{d}^dx \,\, r^{-\tilde{d}}  \left\lbrace \phi^\ast \phi - \frac{\phi^2_B}{2} \right\rbrace.
\end{eqnarray}
Using it to define the renormalized action for the scalar part, we find (neglecting terms of zeroth order in the fluctuations, which do not concern us here)
\begin{eqnarray}
S^\phi_{ren}  & \equiv & \underset{\epsilon\rightarrow 0}{\lim} \left(  S^\phi_{reg} -  S^\phi_{ct}   \right) \nonumber \\
&=& \nu \int \mathrm{d}^dx \,\, \left\lbrace   2 v {\rho}_0 +\rho_0\,\tilde{\rho}_0+   \pi_0 \, \tilde{\pi}_0 \right\rbrace   \ .
\end{eqnarray}
Then the overall variation reads
\begin{eqnarray}
\delta  S^\phi_{ren} & = & \underset{\epsilon\rightarrow 0}{\lim} \left( \delta  S^\phi_{reg} - \delta  S^\phi_{ct}   \right) \nonumber \\
&=& 2 \nu \int \mathrm{d}^dx \,\, \left\lbrace \delta  \rho_0 \left( \tilde{\rho}_0 + v \right)  + \delta  \pi_0 \, \tilde{\pi}_0 \right\rbrace \ ,
\end{eqnarray}
showing explicitly that our counterterm selects $\rho_0$ and $\pi_0$ to play the role of the sources, i.e. their variations have to vanish on the boundary $r=0$ to satisfy the variational principle.

For the transverse sector renormalization, it is again exactly as in \cite{Argurio:2017irz}, to which we refer for the details.
Suffice here to state the only relevant piece in the renormalized action
\begin{eqnarray}
S^T_{ren}
& = & \int \mathrm{d}^dx \,\, \left(\tilde{d}/2 - 1\right) \left(t_0\right)_i\left( \tilde{t}_0 \right)_i\ , 
\end{eqnarray}
up to possible local terms when $\tilde{d}$ is even and strictly bigger than 4. Considering the variation, we find that $t_0$ is identified with the source, as expected.

We finally consider the renormalization of the temporal and longitudinal sectors.
We will treat the case $d=z+1$ (i.e. $\tilde{d} =2z$) in detail 
and see how the result is generalized to any $d$ and $z$ satisfying $d < z+1$.

When $d = z+1$, the expansions for the temporal and longitudinal modes until subleading order reads\footnote{Note that in general, the equation of motion (\ref{eqAL}) leads to a simplification for the expansion of $L$, setting to zero all the possible coefficients between $l_0$ and $\tilde{l}_{0}$, and 
without logarithms for any $\tilde{d}$.}
\begin{eqnarray}
A_t  & \overset{r \rightarrow 0 }{\sim} & \tilde{a}_0  \, \ln r + a_0 + \cdots    \label{expAt1} \\
L & \overset{r \rightarrow 0 }{\sim} & l_0 +  \tilde{l}_{0}  \, r^{2(z-1)}  + \cdots   \label{expAL1}
\end{eqnarray}
It leads to
\begin{eqnarray}
 S_{reg}^{t/ L} &=&   \underset{r=\epsilon}{\int} \mathrm{d}^{d}x \,\, \Big\lbrace -   (z-1)  l_0 \,\partial_j^2 \,   \tilde{l}_0 - \frac{\kappa}{2}  \, \tilde{a}_0 \, \tilde{a}_0\, \ln r  -  \frac{\kappa}{2}  \, a_0 \, \tilde{a}_0 + \cdots   \Big\rbrace
\end{eqnarray}
and, for the variation
\begin{eqnarray}
\delta  S_{reg}^{t/ L} &=&   \underset{r=\epsilon}{\int} \mathrm{d}^{d}x \,\, \Big\lbrace -  2 (z-1) \delta  l_0 \,\partial_j^2 \,   \tilde{l}_0 - \kappa \,   \delta  \tilde{a}_0 \, \tilde{a}_0\, \ln r  - \kappa \,   \delta  a_0 \, \tilde{a}_0 + \cdots   \Big\rbrace\ .
\end{eqnarray}
We see that the only divergence is logarithmic and takes place for the temporal component of the vector. 

As in \cite{Argurio:2016xih}, we explore now two ways of renormalizing this sector.
Adding a mass-like counterterm
\begin{eqnarray}
\tilde{S}^{t/ L}_{ct} & \equiv & - \kappa \underset{r=\epsilon}{\int} \mathrm{d}^{d}x \,\, \frac{\left(A_t - \partial_t L\right)^2}{2 \ln r} \label{CT0}
\end{eqnarray}
gives the following renormalized expression for the variation 
\begin{eqnarray}
\delta  \tilde{S}^{t/ L}_{ren} &=&  \underset{r=\epsilon}{\int} \mathrm{d}^{d}x \,\,  \Big\lbrace -  2 (z-1) \delta  l_0 \,\partial_i^2 \,   \tilde{l}_0 - \kappa \,  \tilde{a}_0\, \partial_t\delta  l_0  + \kappa \,   \delta \tilde{a}_0 \left( a_0 - \partial_t l_0 \right)   \Big\rbrace.
\end{eqnarray}
which exhibits $\tilde{a}_0$ and $l_0$ in the role of the sources. This choice, which we can call ordinary quantization, is not good since $\tilde{a}_0$ does not transform under the residual gauge transformation. Hence, it cannot reproduce a source for $\mathcal{J}_t$ on the QFT side of the correspondence if $\mathcal{J}_\mu$ is a  conserved current.

Inspired again by \cite{Argurio:2016xih}, we propose the following counterterm\footnote{See also \cite{Cvetic:2016eiv,Erdmenger:2016jjg} for similar counterterms, in different set-ups.}
\begin{eqnarray}
S^{t}_{ct} & \equiv & - \kappa \underset{r=\epsilon}{\int} \mathrm{d}^{d}x \,\, \frac{\ln r}{2}\left(r \partial_r A_t \right)^2\ . \label{CT1}
\end{eqnarray}
We note that it can be obtained by adding a term of the Legendre transform kind to (\ref{CT0}):
\begin{eqnarray}
\lim_{\epsilon \rightarrow 0} S^{t}_{ct}  &=& \lim_{\epsilon \rightarrow 0}  \left( -\kappa \underset{r=\epsilon}{\int} \mathrm{d}^dx\, \Big\lbrace (A_t - \partial_t L) \, r\partial_r A_t\Big\rbrace - \tilde{S}^{t/ L}_{ct}\right)\ .
\end{eqnarray}
We find
\begin{eqnarray}
S^{t/ L}_{ren} & \equiv &  \lim_{\epsilon \rightarrow 0} \left( S^{t/ L}_{reg} - S^{t}_{ct} \right) \nonumber \\
& = & {\int} \mathrm{d}^{d}x \,\, \Big\lbrace -   (z-1)  l_0 \,\partial_j^2 \,   \tilde{l}_0 -  \frac{\kappa}{2}  \, a_0 \, \tilde{a}_0    \Big\rbrace\ ,
\end{eqnarray}
and the expression for the variation
\begin{eqnarray}
\delta S^{t/ L}_{ren} &=&  \underset{r=\epsilon}{\int} \mathrm{d}^{d}x \,\,  \Big\lbrace -  2 (z-1) \delta  l_0 \,\partial_j^2 \,   \tilde{l}_0 - \kappa \,   \delta  a_0 \, \tilde{a}_0   \Big\rbrace\ ,
\end{eqnarray}
which is consistent with $a_0$ having the correct gauge transformation for being the source of the temporal component of a conserved current.
Since the source is the subleading term in the expansion, we see that we have to choose ``alternative quantization'' \cite{Klebanov:1999tb} for the bulk field $A_t$, and just for it.

Using the constraint (\ref{eqC}), $\tilde{l}_0$ can be expressed in terms of other coefficients 
\begin{eqnarray}
- \kappa \partial_t \tilde{a}_0 + (2z-2) \partial_j^2 \tilde{l}_0-2\nu w  \tilde{\pi}_0 + 2 \nu v \pi_0 &=& 0.
\end{eqnarray}
Plugging it inside our renormalized action, we find
\begin{eqnarray}
S^{t/ L}_{ren}& = & {\int} \mathrm{d}^{d}x \,\, \Big\lbrace - \frac{\kappa}{2}  l_0 \,\partial_t \tilde{a}_0 - \nu \, l_0\left( w \tilde{\pi}_0 - v\pi_0 \right) -  \frac{\kappa}{2}  \, a_0 \, \tilde{a}_0    \Big\rbrace \nonumber \\
 &= & {\int} \mathrm{d}^{d}x \,\, \Big\lbrace - \frac{\kappa}{2} \left( a_0 - \partial_t l_0 \right) \tilde{a}_0 - \nu \, l_0\left( w \tilde{\pi}_0 - v\pi_0 \right)   \Big\rbrace\ .
\end{eqnarray}

Generalizing now to the case $d<z+1$, the near boundary expansions of the bulk fields remain the same except for the temporal sector, where it is given by \eqref{expAt}
\begin{eqnarray}
A_t  & \overset{r \rightarrow 0 }{\sim} & \tilde{a}_0  \, r^{-(2z-\tilde{d})} + \cdots + a_0 + \cdots
\end{eqnarray}
with possibly also a $\ln r$ term if $z-\tilde{d}/2$ is a positive integer.

As in the case $d=z+1$, the longitudinal sector will not bring any divergence. Hence, we focus on the variation of the temporal part, whose relevant terms are 
\begin{eqnarray}
\delta  S_{reg}^{t} & = &  \kappa(2z-\tilde{d}) \underset{r=\epsilon}{\int} \mathrm{d}^{d}x \,\, \Big\lbrace   \delta \tilde{a}_0 \, \tilde{a}_0 r^{-(2z-\tilde{d})} 
+ \cdots + \delta a_0\, \tilde{a}_0 + \cdots \Big\rbrace \label{varSt}
\end{eqnarray}
We directly go to alternative quantization to see if an adapted version of the counterterm (\ref{CT1}) remains a good choice. The numerical coefficient is fixed to cancel the hardest divergence of the regularized action. We propose
\begin{eqnarray}
 S^{t}_{ct}  & \equiv  &  \frac{\kappa}{2}\underset{r=\epsilon}{\int} \mathrm{d}^d x \,\, r^{2z-\tilde{d}} \frac{\left( r \partial_rA_t \right)^2 }{2z-\tilde{d}}\ .
\end{eqnarray}
Note that this term carries the correct power of $r$ to be covariantly defined with respect to the metric near the boundary.
If $2z-\tilde{d}>2$, we will also need to introduce further counterterms of the same kind as the one above to compensate all subleading  divergences
\begin{eqnarray}
S^{t}_{ct(k)}  & \equiv  & - \frac{\kappa}{2}\underset{r=\epsilon}{\int} \mathrm{d}^d x \,\, r^{2z-\tilde{d} +2k} \frac{\left( r \partial_rA_t \right) \partial_j^{2k} \left( r \partial_r A_t \right)}{c_k} \label{SCTtk}
\end{eqnarray}
with $k$ positive integers and $c_k$ numerical coefficients that are straightforward to determine. 
None of the counterterms will affect the finite term proportional to $\tilde{a}_0$ in $S_{reg}^{t}$. As a consequence, $a_0$ remains a source of the renormalized action, as (\ref{varSt}) is pointing. Putting temporal and longitudinal pieces together, we find
\begin{eqnarray}
 S_{ren}^{t/L} 
& = &   {\int} \mathrm{d}^{d}x \,\, \left\lbrace\kappa (z-\tilde{d}/2)\, a_0\, \tilde{a}_0 - (\tilde{d}/2 - 1) \, l_0 \partial^2_j \tilde{l}_0\right\rbrace\ ,
\end{eqnarray}
and for the variation
\begin{eqnarray}
 \delta  S_{ren}^{t/L} &=&   {\int} \mathrm{d}^{d}x \,\,  \left\lbrace\kappa (2z-\tilde{d})\, \delta a_0\, \tilde{a}_0 - (\tilde{d} -2) \, \delta l_0 \partial^2_j \tilde{l}_0\right\rbrace\ .
 \end{eqnarray}
To get rid of $\tilde{l}_0$, we use again the constraint (\ref{eqC}). Its first order now gives 
\begin{eqnarray}
  \kappa (2z - \tilde{d})\partial_t \tilde{a}_0 + (\tilde{d}-2)  \partial_j^2 \tilde{l}_0-2\nu w  \tilde{\pi}_0 + 2 \nu v \pi_0 &=& 0.
\end{eqnarray}
Then, we find
\begin{eqnarray}
S_{ren}^{t/L}\left[a_0, l_0, \pi_0\right] & = &  {\int} \mathrm{d}^{d}x \,\,  \left\lbrace  \kappa (z-\tilde{d}/2) \left(  a_0 - \partial_t l_0 \right) \tilde{a}_0  - \nu\, l_0 \left( w \tilde{\pi}_0 - v \pi_0 \right)\right\rbrace \ ,
\end{eqnarray}
up to possible local terms if $z- \tilde{d}/2$ is a positive integer.

We can summarize our results for all $d \leqslant z+1$ into the expression 
\begin{eqnarray}
S_{ren}^{t/L}& = &  {\int} \mathrm{d}^{d}x \,\,  \left\lbrace \frac{ \bar{\kappa}}{2} \left(  a_0 - \partial_t l_0 \right) \tilde{a}_0  - \nu\, l_0 \left( w \tilde{\pi}_0 - v \pi_0 \right)\right\rbrace 
\end{eqnarray}
with
\begin{equation}
 \bar{\kappa} \hspace{0.2cm} \equiv  \hspace{0.2cm} \left\{
\begin{array}{ll}
- \kappa &  \text{if} \,\,\, d =z+1 \ ,\\
\kappa(2z - \tilde{d}) &  \text{if} \,\,\, d <z+1 \ .
\end{array}
\right.
\end{equation}

The sum of the renormalized actions for every sector gives the complete gauge invariant effective action that can be used to define the partition function of the QFT
\begin{eqnarray}
S_{ren} & \equiv & S_{ren}^T + S_{ren}^{t/L}+ S_{ren}^\phi\ .
\end{eqnarray}
Thus, we find
\begin{eqnarray}
 S_{ren} \left[t_0,a_0 ,l_0,\rho_0,\pi_0\right] &=& \frac{1}{2}  \int \mathrm{d}^dx \,\, \Big\lbrace  (\tilde{d} -2 ) (t_0)_i(\tilde{t}_0)_i + \bar{\kappa}  \left( a_0 - \partial_t l_0 \right)\tilde{a}_0  \\ 
 & & + 2\nu\, \Big( \rho_0\tilde{\rho}_0 +  \left(\pi_0 - w l_0\right)\left( \tilde{\pi}_0 - v l_0\right) +  v \, \left( 2\rho_0 + 2\pi_0l_0- w l_0 l_0 \right))  \Big)   \Big\rbrace \ .\nonumber
\end{eqnarray}
The equations of motion for the fluctuations relate, through the deep bulk (IR) boundary conditions, the gauge invariant combinations of the tilded coefficients to the gauge invariant combinations of the sources by non-local operators 
\begin{eqnarray}
\tilde{a}_0 & = & \mathfrak{F}_a \,  \left( a_0 - \partial_t l_0\right) + \mathfrak{F}_\pi \,  \left( \pi_0 - w l_0\right)\ , \\
\tilde{\rho}_0 & = &  \mathfrak{G}_\rho \,   \rho_0 \ , \\
\tilde{\pi}_0 - v l_0 & = & \mathfrak{H}_a \,   \left( a_0 - \partial_t l_0\right)  + \mathfrak{H}_\pi \,   \left( \pi_0 - w l_0\right)\ , \\
(\tilde{t}_0)_i & = & \mathfrak{I}_{t}  \, (t_0)_i \ ,
\end{eqnarray}
where all these operators are non-polynomial functions of the derivatives $\partial_t $ and $\partial_i^2$, and we have taken into account that the transverse and $\rho$ sectors are decoupled.
 
We can thus finally write the renormalized action taking this into account
\begin{eqnarray}
S_{ren} \left[t_0,a_0 ,l_0,\rho_0,\pi_0\right] &=& \frac{1}{2}  \int \mathrm{d}^dx \,\, \Big\lbrace  (\tilde{d} -2) (t_0)_i\, \mathfrak{I}_{t}  \, (t_0)_i \nonumber \\
& & + \bar{\kappa}  \left( a_0 - \partial_t l_0 \right)\Big( \mathfrak{F}_a \,  \left( a_0 - \partial_t l_0\right) +\mathfrak{F}_\pi \,  \left( \pi_0 - w l_0\right) \Big) \nonumber \\ 
& & + 2\nu\, \rho_0 \mathfrak{G}_\rho \,   \rho_0    + 2\nu \,  \left(\pi_0 - w l_0\right)\Big( \mathfrak{H}_a \,   \left( a_0 - \partial_t l_0\right) + \mathfrak{H}_\pi \,   \left( \pi_0 - w l_0\right)\Big)   \nonumber \\
& &  + 2\nu v \, \Big( 2\rho_0 + 2\pi_0l_0- w l_0 l_0 \Big) \Big\rbrace \ .
\end{eqnarray}

Now, considering
\begin{eqnarray}
S_{QFT} & \supset  &  \int \mathrm{d}^dx \,\, \Big\lbrace (t_{0})_i\mathcal{J}^T_i - l_0 \partial_i \mathcal{J}_i  -  a_0 \mathcal{J}_t + \rho_0 \mathrm{Re} \mathcal{O} + \pi_0 \mathrm{Im} \mathcal{O}\Big\rbrace,
\end{eqnarray}
and the holographic correspondence, we can write for example
\begin{eqnarray}
\left< \mathrm{Re}\mathcal{O} (x) \right> &= & \frac{\delta  i S_{ren}}{\delta  i \rho_0(x)}\ ,
\end{eqnarray}
or
\begin{eqnarray}
\left< \mathrm{Im}\mathcal{O}(x) \partial_i \mathcal{J}_i(y)\right> &= & \frac{\delta ^2 i S_{ren}}{\delta  i \pi_0(x)\delta  (-i l_0(y))}\ .
\end{eqnarray}
In this way, we find
\begin{eqnarray}
\left< \mathrm{Re}\mathcal{O} (x)\right> &=  & 2 v \nu, \\
\left< \mathrm{Im}\mathcal{O} (x)\mathrm{Im}\mathcal{O} (0)\right> &= & - i 2\nu  \mathfrak{H}_\pi \delta^d(x), \\
- \left< \mathrm{Im}\mathcal{O}(x) \partial_t \mathcal{J}_t (0)\right> + \left< \mathrm{Im}\mathcal{O} (x) \partial_i \mathcal{J}_i (0)\right>  &= & \left(- i 2 w \nu  \mathfrak{H}_\pi  + 2 i v \nu \right)\delta^d(x)\ .
\end{eqnarray}
The latter relation can be reexpressed as 
\begin{equation}
- \left< \mathrm{Im}\mathcal{O}(x) \partial_t \mathcal{J}_t (0)\right> + \left< \mathrm{Im}\mathcal{O} (x) \partial_i \mathcal{J}_i (0)\right>  =   w \left< \mathrm{Im}\mathcal{O}(x) \mathrm{Im}\mathcal{O}(0) \right>  +  i \left< \mathrm{Re}\mathcal{O} \right> \delta^d(x)\ ,
\end{equation}
which are the Ward identities for a current associated to a symmetry which is broken both spontaneously (by $v$) and explicitly (by $w$).

In the purely spontaneous case, the Ward identities imply the presence of a gapless mode, i.e.~a Goldstone boson.
What our holographic analysis has shown is that the procedure of holographic renormalization is still consistent with the presence of a non-zero VEV $v$. This then indicates that spontaneous symmetry breaking is indeed possible in holographically realized Lifshitz theories in $d\leq z+1$.

\section{Discussion and outlook}
\label{secDISC}
In this paper we have analyzed the possibility to have spontaneous symmetry breaking in theories with Lifshitz scaling, depending on the dimensionality of space-time. First, we considered the issue from the purely field theoretic perspective, and found the expected result: when the mass dimension of a scalar is zero or negative, i.e.~when $d\leq z+1$, large quantum fluctuations in the massless case erase any possibility of having an order, i.e.~a VEV. We then proceeded to consider the same situation in a holographic set-up, suitable for a large $N$ theory. We found there that there is no consistency problem in having a non-zero VEV,\footnote{For instance, in principle a legitimate alternative result could have been to find that it was impossible to cancel all divergencies for $v\neq 0$.} and hence a propagating massless scalar. This is consistent with the expectation that order can be restored in the $N\to\infty$ limit.

With respect to the previous analysis of the relativistic case in \cite{Argurio:2016xih}, we have seen that also in the present case we have to resort to alternative quantization for the vector. However, and this is a novel feature, only the temporal component of the vector has to be treated in this way. Actually, it is the expected gauge symmetry of the renormalized action that ultimately dictates to us this asymmetric treatment of the temporal and spatial components of the bulk vector.\footnote{Note that  a precondition to have a situation opposite to the one that we described, i.e.~alternative quantization only for the spatial components of the vector and ordinary quantizaton for the temporal component, is to have $2z<\tilde{d}<2$, i.e.~$z<1$. We can thus conclude that this possibility does not arise in physically sensible set-ups \cite{Hoyos:2010at}.
}

We now comment on some issues that we did not address in the present paper, but that could be worth investigating.
\begin{itemize}
\item 
In the present paper, we have focused on theories with time-reversal invariance. An obvious generalization is to theories with no such invariance, i.e.~including a term linear in the time derivative. Note that the holographic treatment of \cite{Argurio:2017irz} includes such a case. However, consider a candidate Goldstone mode which has an EFT with Lifshitz scaling and a term linear in $\partial_t$. The latter will be the most relevant kinetic term at lowest energies. We can thus consider a theory where the kinetic term is purely linear. The dimension of the scalar is then $(d-1)/2$, and it is always positive in our case since $d\geq2$. We thus naively do not expect to find any space-time dimension in which spontaneous symmetry breaking is prevented by large vacuum fluctuations. It would be nevertheless interesting to analyze in more details how this works for low dimensions. Also, the same should be true more specifically for type B Goldstone modes \cite{Watanabe:2012hr,Kapustin:2012cr,Argurio:2015via}, which enjoy a kinetic term linear in $\partial_t$. 
\item 
Having shown in this paper that the holographic approach, being pertinent to the $N\to\infty$ limit, allows for spontaneous symmetry breaking, one can ask whether $1/N$ corrections can spoil this result and set the VEV to zero when $d\leq z+1$. This amounts to computing corrections at leading order in the bulk interactions. Eventually, one is led to perform a one-loop integral in all similar to the one performed in section \ref{secQFT}. This approach was followed in \cite{Anninos:2010sq} for the case of $d=3$ and finite temperature, finding that indeed large fluctuations erase the bulk scalar profile dual to the VEV. We expect a similar result also in the cases considered in the present paper. 
\item 
Further, we can ask what happens when temperature is turned on. On the QFT side, a general argument like in \cite{Ma:1974tp} from thermal field theory (see for instance \cite{Bellac:2011kqa}) gives for a massless mode at finite temperature $T=1/\beta$
\begin{equation}
\left< \theta(0,\vec{x})\theta(0) \right>_T  \propto \int \mathrm{d}^{d-1} p \, \frac{e^{i\vec{p}\cdot\vec{x}}}{p^{z}}\left( 1+\frac{2}{e^{\beta p^{z}}-1}\right)\sim 2T \int \mathrm{d}^{d-1} p \, \frac{e^{i\vec{p}\cdot\vec{x}}}{p^{2z}} + \dots\ ,
\end{equation}
where in the last step we have isolated the most IR divergent term. From the latter, we observe that at $T>0$, such integral is generically IR divergent when $d\leq 2z+1$, hence increasing the critical dimension below which spontaneous breaking of continuous symmetries is prevented. Note that for $z=1$ we recover the Mermin-Wagner-Hohenberg theorem \cite{Mermin:1966fe,Hohenberg:1967zz}.\footnote{Without time-reversal invariance, a similar argument would suggest that the critical dimension is now $d\leq z+1$. These theories do not seem to respect the Mermin-Wagner-Hohenberg theorem, in the same way as they do not with the Coleman theorem at zero temperature.
}
In holography, one should study scalar profiles in Lifshitz black hole spacetimes (see e.g.~\cite{Danielsson:2009gi,Bertoldi:2009vn,Balasubramanian:2009rx,Tarrio:2011de,Korovin:2013nha}). In the latter set-up, one does not expect any variation with respect to our results if the spacetime metric is asymptotic to the pure Lifshitz one. Bulk $1/N$ corrections should on the other hand be sensitive to the presence of the black hole horizon.
\item 
Finally, it would be interesting to explore possible realistic systems which display Lifshitz scaling (see \cite{Hartnoll:2009sz} and references therein), in the $d\leq z+1$ regime, to verify that indeed the spontaneous breaking of continuous symmetries does not take place. That would apply to systems in two spatial dimensions with $z\geq2$, or in three spatial dimensions with $z\geq3$. Finding such systems could open the way to an experimental verification of the phenomenon discussed in this paper. 
\end{itemize}

\section*{Acknowledgments} 
We would like to thank Yegor Korovin, Andrea Marzolla, Daniele Musso and Pierluigi Niro for helpful discussions and feedback.
We acknowledge support by IISN-Belgium (convention 4.4503.15) and by the F.R.S.-FNRS under the `Excellence of Science' EOS be.h project n.~30820817. R.A. is a Research Director of the Fonds de la Recherche Scientifique-F.N.R.S. (Belgium). A.P. is a FRIA grantee of the Fonds de la Recherche Scientifique-F.N.R.S. (Belgium).

\end{document}